\documentclass[sn-standardnature, referee]{sn-jnl} 

\usepackage{gensymb}    
\usepackage{comment}    
\usepackage{wasysym}    
\usepackage{ulem}       
\usepackage{multirow}   
\usepackage{color}
\usepackage[T1]{fontenc}

\jyear{2025}%
\unnumbered 

\begin{document}

\title[Forming Mercury]{Forming Mercury by a grazing giant collision involving similar mass bodies}

\author*[1,4]{\fnm{Patrick} \sur{Franco}}\email{patrickoliveira@on.br}
\author[1]{\fnm{Fernando} \sur{Roig}}\email{froig@on.br} 
\author[2]{\fnm{Othon C.} \sur{Winter}}\email{othon.winter@unesp.br} 
\author[2,3,4]{\fnm{Rafael} \sur{Sfair}}\email{rafael.sfair@unesp.br} 
\author[3]{\fnm{Christoph} \sur{Burger}}\email{christoph.burger@uni-tuebingen.de}
\author[3]{\fnm{Christoph M.} \sur{Schäfer}}\email{ch.schaefer@uni-tuebingen.de}
\affil*[1]{\orgdiv{Department of Astronomy}, \orgname{National Observatory}, \orgaddress{\street{Rua Gal. José Cristino, 77}, \city{Rio de Janeiro}, \postcode{20921-400}, \state{RJ}, \country{Brazil}}}
\affil[2]{\orgdiv{Grupo de Din\^ amica Orbital e Planetologia}, \orgname{São Paulo State University}, \orgaddress{\street{Av. Dr. Ariberto Pereira da Cunha, 333}, \city{Guaratinguetá}, \postcode{12516-410}, \state{SP}, \country{Brazil}}}
\affil[3]{\orgdiv{Institute of Astronomy and Astrophysics}, \orgname{University of T{\"u}bingen}, \orgaddress{\street{Auf der Morgenstelle 10}, \postcode{72076} \city{T{\"u}bingen}, \country{Germany}}}
\affil[4]{\orgdiv{CAGE}, \orgname{Institut de physique du globe de Paris, Université Paris Cité, CNR}, \orgaddress{\street{1 rue Jussieu}, \postcode{F-75005}, \city{Paris}, \country{France}}}
\affil[5]{\orgdiv{LESIA}, \orgname{Observatoire de Paris, Université PSL, CNRS}, \orgaddress{\street{Sorbonne Université, 5 place Jules Janssen}, \postcode{92190}, \city{Meudon}, \country{France}}}

\abstract{
The origin of Mercury still remains poorly understood compared to the other rocky planets of the Solar System. One of the most relevant constraints that any formation model has to fulfill refers to its internal structure, with a predominant iron core covered by a thin silicate layer. This led to the idea that it could be the product of a mantle stripping caused by a giant impact. Previous studies in this line focused on binary collisions involving bodies of very different masses. However, such collisions are actually rare in N-body simulations of terrestrial planet formation, whereas collisions involving similar mass bodies appear to be more frequent. Here, we perform smooth particle hydrodynamics simulations to investigate the conditions under which collisions of similar mass bodies are able to form a Mercury-like planet. Our results show that such collisions can fulfill the necessary constraints in terms of mass (0.055 $M_\oplus$) and composition (30/70 silicate-to-iron mass ratio) within less than 5\%, as long as the impact angles and velocities are properly adjusted
according to well established scaling laws.}


\maketitle


\section*{Main}

Hypotheses about the formation of the planet Mercury have become even more conflicting since the gamma-ray spectroscopic observations of the \textit{MESSENGER} mission \citep[][]{Peplowski11, Nittleretal2011, Evans15}. In addition to the long-investigated high uncompressed density of Mercury \citep[][]{Benz88, Asphaug_Agnor_04, Benz2007, Asphaug14, ebel_stewart18}, due to its very large iron core mass fraction of $\sim$70\%, \textit{MESSENGER} revealed that Mercury's surface is not depleted of moderately volatile elements, despite its thin mantle layer. Reconciling both the geophysical structure and volatile composition of Mercury simultaneously, within a unified formation model, constitutes the so-called ``Mercury problem'', which has proven to be extremely challenging \citep{ebel_stewart18}.

Reproducing Mercury's mass also remains a major challenge for studies based on N-body numerical simulations, which attempt to match as many inner solar system constraints as possible \citep[][]{Raymondetal20}. 
In such simulations, few bodies form in the region currently occupied by Mercury, and those that do form are often more massive than Mercury \citep[][]{R09, Izidoroetal15,LI17,IzidoroRaymond18, Clement19b, francoetal22, Clement-etal-2023}. This is in part due to the assumed initial mass distributions in the simulated protoplanetary discs and partly also due to the treatment of collisions as perfectly inelastic mergers \citep[][]{Duncan1998,Chambers1999}. However, even in simulations where collisional fragmentation and associated mass loss are accounted for through the application of scaling laws, 
the results do not show a significant improvement in terms of the final mass of Mercury-like bodies \citep[][]{Chambers13, Emsenhuberetal20}. The use of hybrid N-body + SPH simulations, in which all collisions are evaluated through individual hydrodynamical simulations \citep{burgeretal20}, partly helps to overcome this issue.

One of the hypotheses proposed to explain Mercury's atypical internal structure is that a substantial portion of its primordial mantle was removed as a result of planetary-scale collisions, either a single decisive event or multiple erosive impacts \citep{Cameronetal1988, Benz88, Benz2007, Asphaug10, Asphaug_Reufer_14, saridetal14, Chau18}. In the \textsl{canonical} giant impact scenario, a single head-on giant impact between a planetary embryo with 2.25 times the current Mercury's mass and an object six times smaller \cite{Benz88,Benz2007}, succeeded in removing the embryo's mantle to resemble present-day Mercury's internal structure. However, in addition to the problem of silicate reaccretion in the aftermath of the collision \citep{Gladman-Coffey-2009},
such an energetic impact scenario requires the projectile to be in an extreme eccentric orbit prior to the event, but it is predicted to be rare in simulations of terrestrial planet formation \citep{Obrien, Asphaug_Reufer_14, Jacksonetal2018, Chau18, stewartetal16, ebel_stewart18, Clement19a}. To avoid these issues, subsequent studies suggested a grazing collision, in the so-called \textsl{hit-and-run} regime, in which the proto-Mercury strikes an Earth-mass body and loses a significant amount of mantle material \citep[][]{Asphaug_Reufer_14, saridetal14, Chau18}.

Alternative hypotheses, which do not invoke giant collisional events, suggest that Mercury already had an iron-rich composition in the final stages of terrestrial planet formation. This would arise from the local condensation of specific chemical elements, such as metallic iron, during the early stages of protoplanetary disc accretion \citep[][]{EbelAlexander2011,Wurmetal2013,Pignataleetal2016,ebel_stewart18,KrussWurm2018,KrussWurm2020}.

Here, we focus on the scenario of hit-and-run collisions. 
Such collisions may account for almost one third of all collisions in numerical simulations of the evolution of protoplanetary discs with N-body codes \citep[][]{SL12,francoetal22}.  
However, recent studies indicate that hit-and-run collisions
involving a target with roughly ten times the mass of the projectile, as suggested by \citep[][]{Asphaug_Reufer_14}, are less likely to occur \citep{Clement19a,Clement19b, burgeretal20, clementetal21, francoetal22}. In particular, no occurrence of this impact arrangement was recorded over 33 different systems, corresponding to $\sim$500 simulations of protoplanetary discs analyzed in \cite{francoetal22}.
Moreover, such collisions would also require extreme eccentric orbits \citep{Jacksonetal2018, Clement19a} to promote a sufficiently energetic impact, as required in the canonical scenario. However, collisions between a planetary embryo and an Earth-sized body generally have relatively low velocities, resulting in a merging event rather than an erosive hit-and-run.

\begin{figure}[h]%
    \centering
    \includegraphics[width=\textwidth]{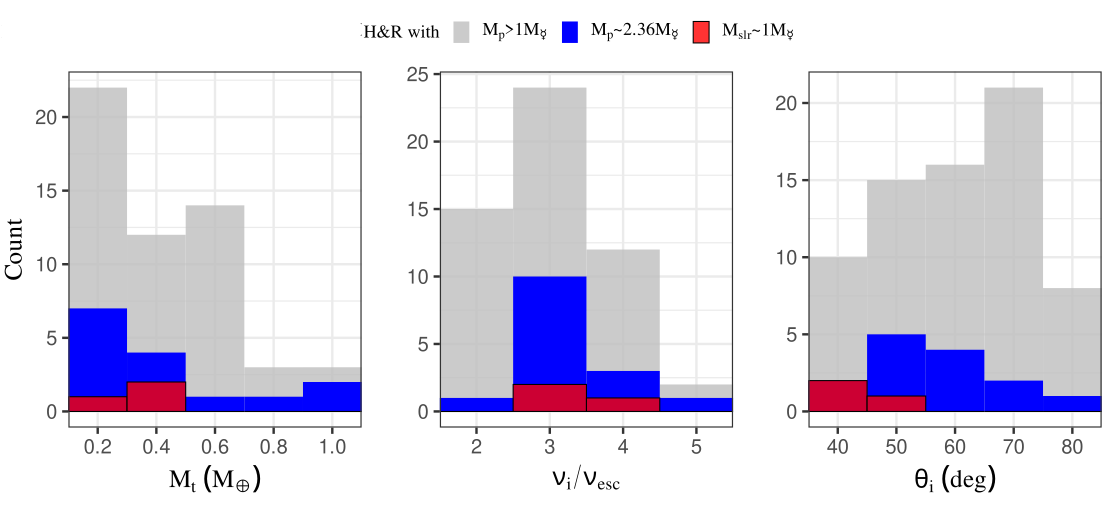}
    \caption{Distribution of hit-and-run collisions detected in N-body simulations of terrestrial planets accretion in the solar system, in terms of target mass (left), scaled impact velocities (middle), and impact angles (right). The results have been compiled from a reanalysis of the eight different configurations of protoplanetary discs analyzed in \cite{francoetal22}. The gray histograms represent all the collisions occurring in the hit-and-run regime where the projectile has at least one Mercury mass ($\sim0.055~M_{\oplus}$). The blue histograms are the subset where the projectile has nearly a Mars-like mass ($\sim0.13~M_{\oplus}$). The red histograms are the sub-subset where the collisions produce a Mercury analog, i. e., a second largest fragment with a mass of $(1\pm0.25)M_{\mercury}$, according to the scaling laws \cite{LS12,SL12}.}
    \label{histo}
\end{figure}

On the other hand, \cite{francoetal22} observed a higher occurrence of grazing giant impacts involving similar mass bodies, in which roughly Mars-sized projectiles collide with targets just 1.5 to 4 times more massive ($\sim20\%$ of all systems of protoplanetary discs, being almost always present in groups with more dynamically excited discs - see Table 5 in the refereed work). A reanalysis of the simulation results from such work shows that although collisions generally have a stochastic character in numerical simulations, hit-and-run impacts are more common under certain parameters, as shown in Fig.~\ref{histo}. However, due to the low frequency of giant impacts ($\sim$1-2\% of all collisions in most of our systems), such collisions are still considered rare events and not observed in all of our protoplanetary systems.

Our goal in the present work is to show that while the hit-and-run scenarios investigated in the past are unlikely occurrences in N-body simulations, the equal-mass low-velocity hit-and-run scenario is more promising in producing bodies with mass and internal structure similar to Mercury's. Then, it is important to highlight that in this study we did not investigate the long-term dynamical evolution of the impact remnants (orbit origin/fate of the colliding/remnant bodies). This evolution would occur in the context of the early Solar System, when the disk still contained a significant population of planetesimals and planetary embryos in regions from 0.5 au to 1.0 au. A full exploration of these processes would require extensive additional investigation and is beyond the scope of this work.

Nevertheless, we assume that this impact could be inserted in the initial tens of millions of years of the final phase of planet formation, when the disc still contains a large population of planetary embryos, as well as the still-growing proto-Venus and proto-Earth. Also, from \cite{francoetal22} data, this impact can preferentially take place in regions ranging from 0.5 au to 1.0 au, rather than Mercury's current orbit. Thus, theoretical constraints on the fate of remnants regarding the current solar system arrangement may be addressed more easily.

The scenario of hit-and-run collisions between similar mass bodies has been partially explored by \citep{saridetal14,Chau18,Gabriel_etal_2020}. The approach in those studies was to consider a parametric space that sweeps several collision scenarios, testing a large number of combinations of impact parameters, such as bodies' masses, relative velocities and impact angles.
Apart from being computationally costly, this approach usually disregards feasible impact setups \cite{Chambers23} to reproduce the Mercury structure in a single giant impact.
Then, more realistic impact geometries for obtaining Mercury analogues remain elusive in these studies, even assuming reasonable mass configurations.

In contrast to those previous approaches, here we
look into the specific setups that are able to correctly reproduce Mercury's internal structure, namely the overall mass and core-mass fraction, and we have set strict success criteria. We estimate the adequate impact parameters, in particular the relative velocities and impact angles, using the well established scaling laws of \citep{LS12} to provide: (i) the lowest possible impact energy for each configuration, thus matching the average collision energy observed in protoplanetary disc simulations, and (ii) the second largest remnant of the collision with a mass similar to that of present-day Mercury. 
We perform a series of smooth particle hydrodynamics (SPH) hit-and-run impact simulations, the technical details of which are described in Methods section.

\begin{figure}[h]%
    \centering
    \includegraphics[width=0.7\textwidth]{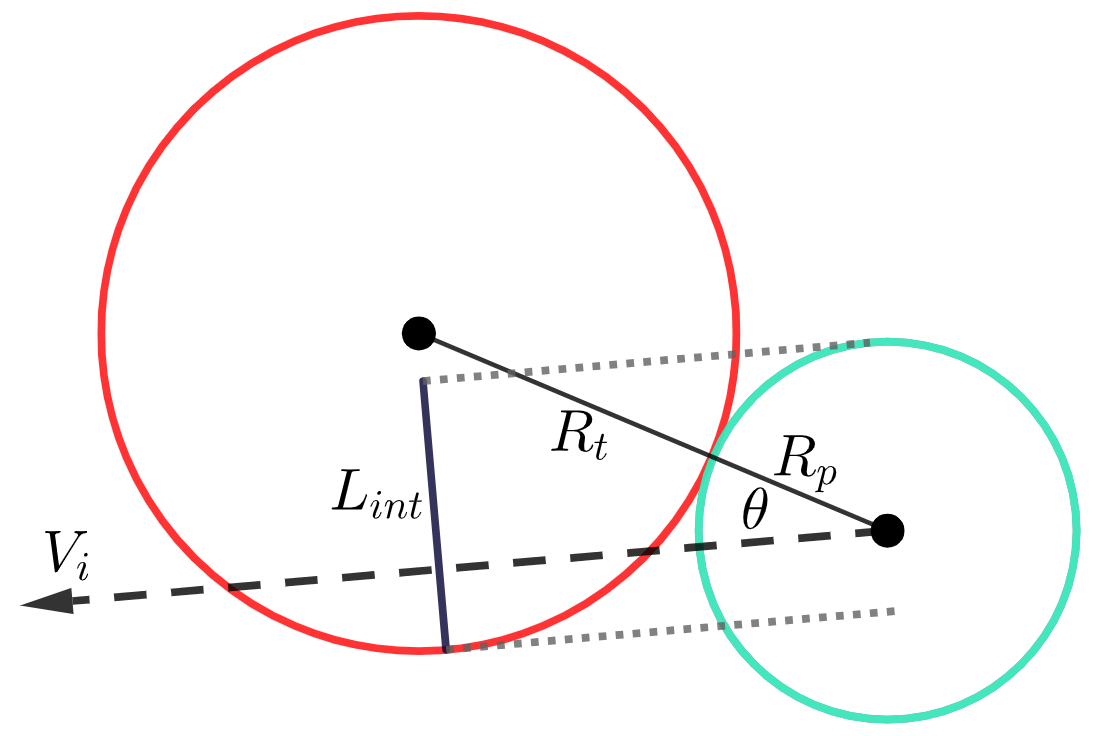}
    \caption{Schematic of a collision configuration. $V_i$ denotes the relative impact velocity between the target of radius $R_t$ and the projectile of radius $R_p$. The impact angle, $\theta$, is defined as the angle between the lines $R_t+R_p$ and the relative velocity vector at the moment of first contact. $L_{\mathrm{int}}$ represents the interacting length, defined as the projected length of the projectile that overlaps the target.}
    \label{col_scheme}
\end{figure}

We consider a proto-Mercury with a mass of $2.36\,M_{\mercury}$ ($\simeq0.13\,M_\oplus$) with an iron-mass fraction of 0.3, which strikes a target with masses ranging from $0.2\,M_{\oplus}$ to $0.6\,M_{\oplus}$ with iron-mass fractions of 0.3 and 0.5, at relatively low velocities (2.8 to 3.8 times the mutual escape velocity, Eq. \ref{equation_vesc}). In particular, the combination of this range of impact velocities involving an impactor with mass exceeding that of current Mercury is present in more than 50\% of all our protoplanetary systems.
The tested impact configurations have been set according to the relationship between the impact angle and the interacting length, $L_\mathrm{int}$, which gives an idea of the degree of overlap between the projectile and the target during the collision (see Methods, Fig. \ref{col_scheme}). For the given projectile-target mass arrangements, we have found some successful configurations of impact angle $\theta$ and velocity $v$, producing a \textsl{second largest remnant} analogue to Mercury in terms of both total mass, $M_\mathrm{slr}$, and iron core-mass ratio, $Z_\mathrm{Fe,slr}$. The results of all our simulations are compiled in Table \ref{configurations}. The simulations have been divided into two groups, according to the initial value of iron-mass fraction of the target: $Z_\mathrm{Fe}=0.5$ for the simulations in group A, and $Z_\mathrm{Fe}=0.3$ for the simulations in group B.

\begin{figure}[h]%
    \centering
    \includegraphics[width=\textwidth]{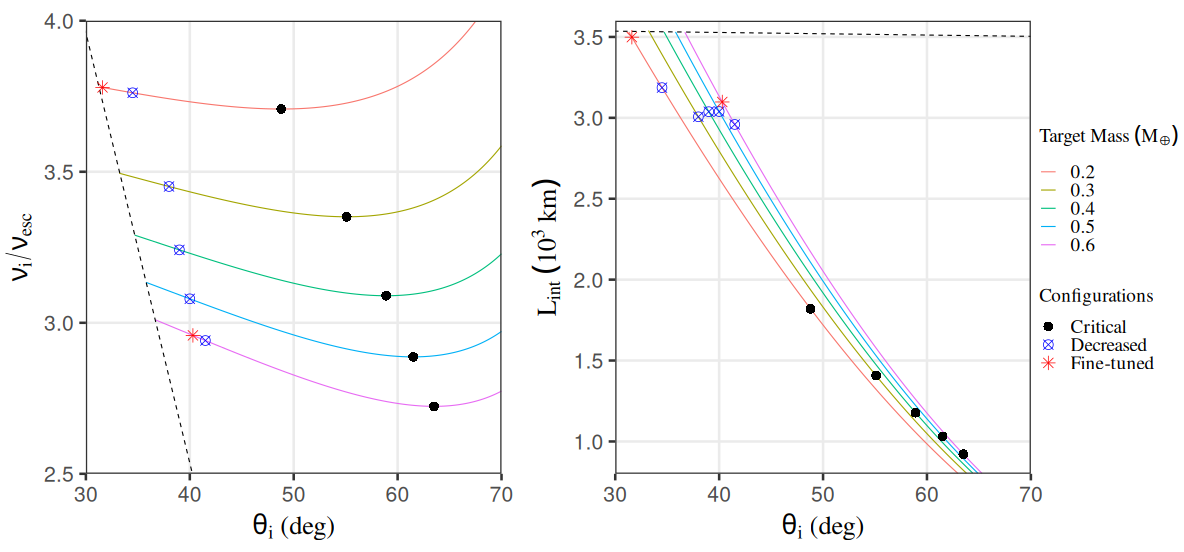}
    \caption{Scaled impact velocity (left panel) and interacting length ($L_{\mathrm{int}}$, right panel) as a function of impact angle for the simulations in the A group configuration. Each curve corresponds to the behavior predicted by the scaling laws \citep[][]{LS12,SL12} for a specific combination of target mass colliding with a projectile of $0.13~M_{\oplus}$. The black dots represent the absolute minimum of each curve (left panel), which correspond to the critical impact angles (AC group). The blue crossed circles represent the set of decreased impact angles (AD group). The red marks are the fine-tuned impact angles (AF group). The dashed line corresponds to the limit between the grazing (hit-and-run) and non-grazing regimes, to the right and left, respectively.}
    \label{fig_vi_thetai}
\end{figure}

In a first series of simulations, we adopted the set of parameters provided by the scaling laws that suggests the smallest impact velocity necessary to form a Mercury-size post-impact remnant in the hit-and-run regime. These impact velocities are correlated with the critical impact angle $\theta_\mathrm{crit}$ (see Methods, Fig. \ref{fig_vi_thetai}). This setup did not produce the desired outcomes, where the resulting projectile remnant remains nearly twice the mass of present-day Mercury, and the iron-mass fraction remained below 0.5.

\begin{figure}[h]%
    \centering
    \includegraphics[width=\textwidth]{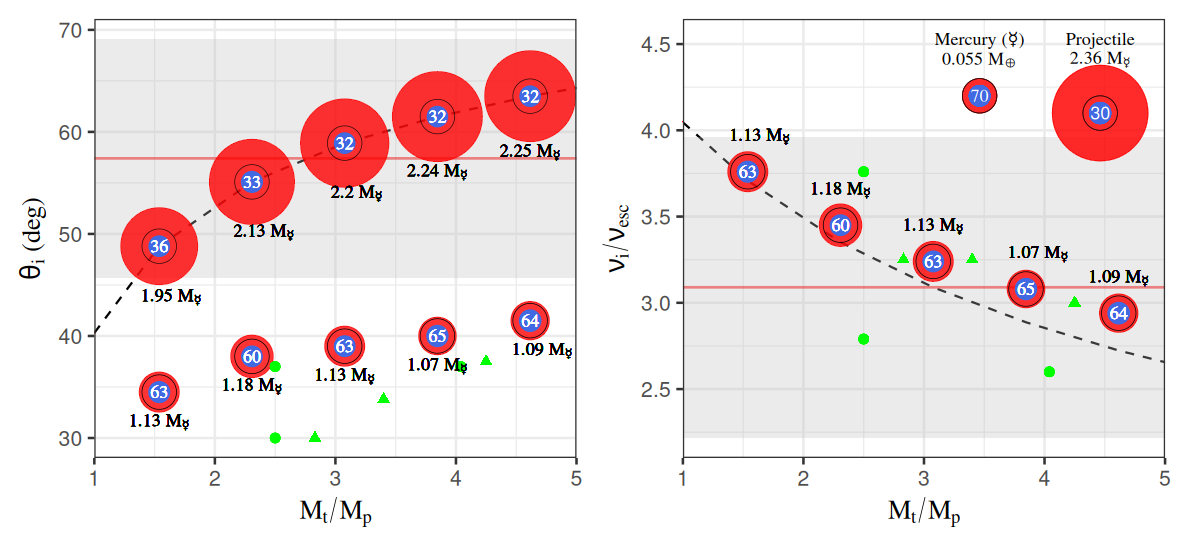}
    \caption{Summary of the outcomes of our SPH simulations for the group A configurations (target core-mass ratio of 0.5). The left panel shows the impact angle, $\theta_i$, as a function of the target mass scaled by the projectile mass for AD and AC groups. The right panel shows the scaled impact velocity, $v_i/v_{\text{esc}}$, also as a function of the target mass scaled by the projectile mass just for AD group to avoid overlapping. Each circle represents the final size and composition (red for silicate, blue for iron) of the second largest remnant. The iron mass fraction (in \%) and the final mass (in $M_{\mercury}$) are labeled for each remnant. The mass and composition of the initial projectile (same for all simulations) and of the present-day Mercury are represented in the top right corner of the right panel. The black dashed curves represent the minimum estimated impact velocity to produce a Mercury-size remnant, according to the scaling laws \citep[][black dots in Fig. \ref{fig_vi_thetai}]{LS12,SL12}. The red solid line is the mean angle and velocity impact, respectively, corresponding to all collisions around current Mercury orbit (from 0.1 to 0.8 au) and the gray areas represent the standard deviation. The green triangles and circles respectively represent roughly the impact parameters of the best results from \cite{Asphaug_Reufer_14} and \cite{Chau18}  (without considering the initial core-mantle ratios of each body in these works). }
    \label{results_vel}
\end{figure}

In a second series of simulations, we reduced the impact angle in order to increase the interacting length, thus producing a more disruptive collision, while maintaining the relationship between $\theta$ and $v$ defined by the scaling laws (see Methods, Fig. \ref{fig_vi_thetai}). We considered values of $\theta$ that were 30\% smaller than the values of $\theta_\mathrm{crit}$, which is equivalent to a $L_{\mathrm{int}}$ of $\sim3000$~km. The general results for these simulations were significantly better than the previous ones, with final $M_\mathrm{slr}\sim1~M_{\mercury}$ and $Z_\mathrm{Fe,slr}\sim0.6$-0.65.
The iron core-mass fractions of the final Mercury analogues resulting from the first and second series of simulations in group A are summarised in Fig. \ref{results_vel}.

\begin{figure}[h]%
    \centering
    \includegraphics[width=\textwidth]{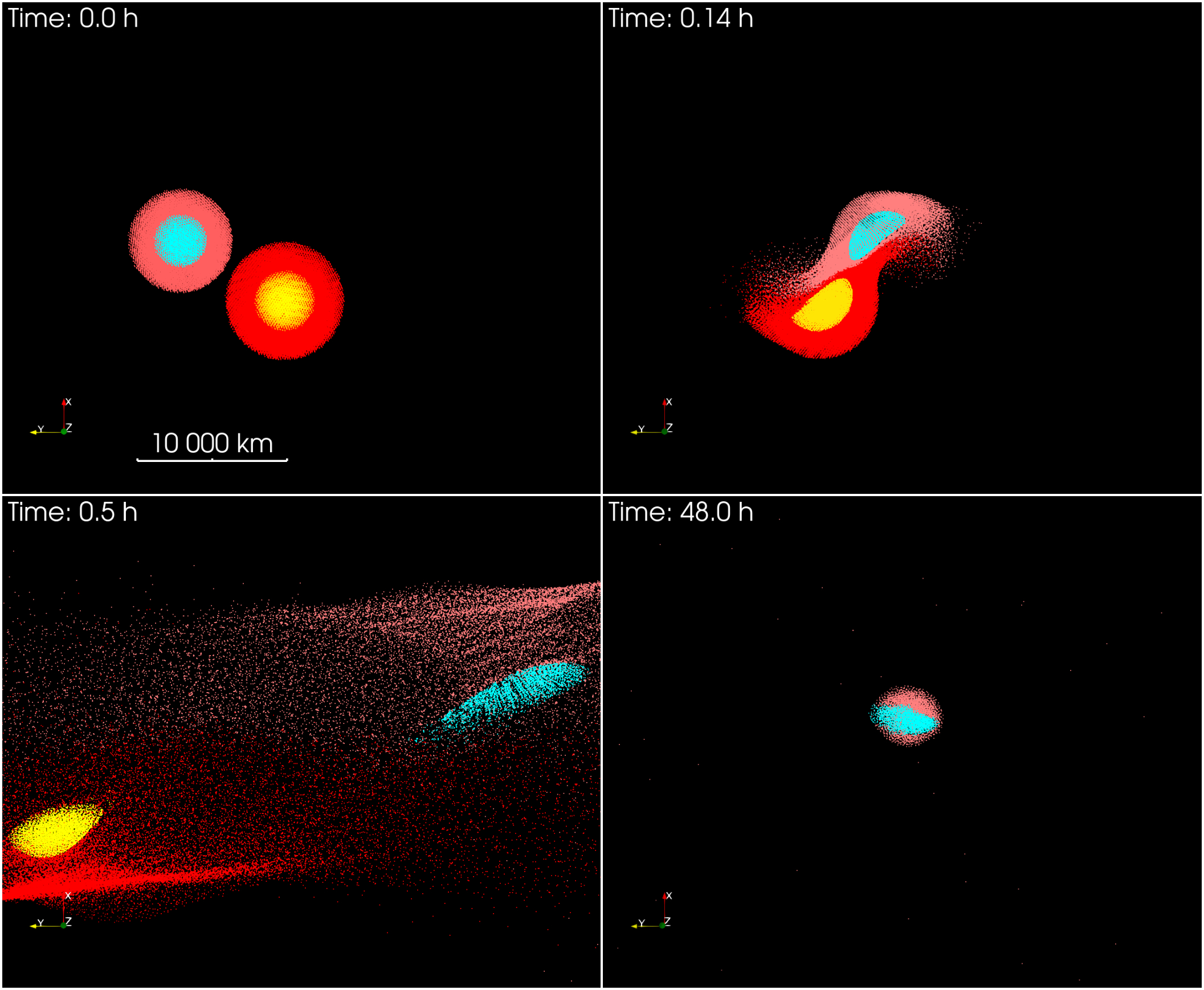}
    \caption{Snapshots of the collision in the BF configuration with a target of $0.2~M_{\oplus}$. The time interval between snapshots is indicated in the top left corner and is not equispaced. The spatial scale and orientation in each snapshot is the same. The proto-Mercury ($0.13~M_{\oplus}$) is represented by a pink mantle and a turquoise core. The target is represented by a red mantle and a yellow core. The impact angle is $32.5\degree$, and the impact velocity is $22.3~\mathrm{km\,s}^{-1}$. The total time span of the simulation is 48 hours. The first frame shows the two bodies with an initial core-mass ratio of 0.3, characterized by a predominantly rocky composition, just before the impact. The final frame showcases the Mercury candidate produced from a relatively low-velocity collision. In the last frame, Mercury candidate ends up with a $Z_{\mathrm{Fe}}$ of 0.68 and mass of $0.056~M_{\oplus}$, values extremely close to those of current Mercury}. 
    
    \label{snap}
\end{figure}

In a third series of simulations, we obtained an even closer match to present-day Mercury, with a slight decrease in impact angles to increase $L_{\mathrm{int}}$ and thus damage, thus improving projectile mantle removal. Given that the higher the mass ratio of the colliding bodies, the greater the damage to the projectile, we set a smaller $L_{\mathrm{int}}$ to balance this damage. Hence, in these simulations we adopted $L_{\mathrm{int}}=3500$~km ($\sim90\%$ of the target radius) for the less massive target ($0.2~M_{\oplus}$), and $L_{\mathrm{int}}=3100~$km ($\sim60\%$ of the target radius) for the more massive target ($0.6~M_{\oplus}$). With this setup, we obtained results in which $M_\mathrm{slr}$ matches the current Mercury mass within 5\%, and $Z_\mathrm{Fe,slr}$ between 0.65-0.75, in good agreement with \cite{Chau18}. Figure \ref{snap} shows four snapshots of the evolution of our best simulation considering a low impact velocity ($\sim 3.8\,v_\mathrm{esc}$). The first snapshot, taken just before impact, shows two bodies with core-mass ratios of 0.3, most rocky-composed, similar to the present terrestrial planets, to enable substantial removal of the mantle. The final snapshot reveals a striking Mercury-like remnant with a total mass and an iron-mass fraction that closely matches Mercury’s current values.

The time scale for the projectile remnant to achieve a mass similar to the present Mercury mass is at least 35 hours after the impact for the simulations in group A, and around 12 hours after the impact for those in group B. In group A, the higher time scale reflects a more pronounced disruption of the projectile due to the higher mean density of the targets compared to those in group B. As a result of the larger damage, the projectile in group A simulations requires more time to gravitationally reaccrete the scattered material until it reaches the same mass as the final projectile in group B simulations.

Our results indicate that a single grazing collision involving two planetary embryos of comparable size successfully produces the present-day Mercury, in terms of mass and iron core-mass fraction.
Not all sets of impact parameters estimated by the scaling laws provide the same results, which is expected since such equations consider bodies of homogeneous density to predict their outcomes. However, impact angles and velocities distributed within narrow ranges of values based on scaling laws predictions (see Fig. \ref{fig_vi_thetai}) provide a second largest remnant with the desired properties in terms of both overall mass and iron-mass fraction. This points to the Mercury-forming impact as a potentially very specific event, very sensitive to small changes of the parameters, especially the impact angle if we consider one single event (see Fig. \ref{results_vel}). In particular, we obtained a Mercury-analogue remnant with impact angles such that the corresponding $L_{\text{int}}$ fell between 3100 and 3500~km, depending on the target size.

Although we have not performed simulations with impact angles larger than critical ones, we do not expect to achieve potential outcomes in such scenarios. The larger the impact angle, the shorter the interacting length between colliding bodies. This configuration requires more impact energy to produce the same result because the damage caused to the projectile decreases as the interacting length is smaller. As a result, the post-impact projectile remains predominantly rocky-composed, as seen in the simulations using the critical angles, where removal of the mantle is insufficient to increase the core-to-mantle fraction of the remnant.

Our simulations show that the mass stripping in the target involved in the collision may be significant, but, in general, its final composition remains relatively unchanged.
The target suffers a more pronounced disruption and a significant modification in composition ($\sim20\%$ and $\sim30\%$ of mass loss for groups A and B, respectively) when the target-to-projectile mass ratio is closer to 1, and even greater disruption when adopting the fine-tuned impact angles that are very close to the grazing regime limit (red marker in Fig. \ref{fig_vi_thetai}).

The scenario of multiple erosive impacts, in which the proto-Mercury is energetically hit by smaller bodies, has also been proposed as a mechanism for its partial removal of the mantle \citep{svetsov11,Chau18}. Reproducing Mercury’s current core-mass fraction requires the cumulative effects of successive erosive impacts by small bodies. The total mass of these impactors must exceed that of proto-Mercury, with impact velocities approaching 30 km/s \citep{svetsov11}. Furthermore, both the material ejected and the planetesimals must avoid being (re)accreted into the proto-Mercury in such events \citep{Chambers13, ebel_stewart18}. The time interval between successive impacts also constrains the model \citep{Chau18}. Such implications affect the feasibility of this scenario. However, there is evidence that small bodies with large core-mass fractions have generally been involved in multiple hit-and-run impacts \citep{burgeretal20,Emsenhuberetal20,Cambionietal21};
therefore, further assessment is needed to determine if this scenario is, in fact, a viable alternative to produce core-dominant terrestrial planets.
 It is important to highlight that, while multiple collisions with smaller impacting embryos are more common than those with similar or larger embryos, the latter scenario is still feasible. \citep{SL12,francoetal22}. Thus, even if mass loss may be caused via multiple small collisions, a giant impact -- likely involving roughly similar sized bodies -- has the potential to erase prior collisional signatures and can essentially imprint Mercury's final composition.

We conclude that a single impact in the hit-and-run regime, widely discussed as the Mercury-forming event, is a viable approach to produce core-dominant bodies by adopting a small range of combined impact parameters (impact velocity and geometry) that do not necessarily require extreme conditions, like overly massive targets. We verified that our results are robust and do not depend on the total time span of SPH simulation or its resolution, considering the final structure of the projectile remnant produced following the impact (see Methods). Future research will explore the long-term dynamic evolution of colliding bodies and determine under what conditions our results remain consistent with Mercury's internal structure while also agreeing with the structure of the inner solar system.

Our study has also predicted a wide range of impact configurations in the hit-and-run regime in such a way that the proto-Mercury either experiences almost no mass loss (at critical or larger impact angles) or sufficient mass loss (at the decreased or fine-tuned impact angles) to reach Mercury's current composition via a single impact. Between these two extremes, it may take more than one grazing giant impact to produce the same result by mantle removal. Thus, a sequence of erosive impacts may also be a viable pathway to building the present-day Mercury. 

In terms of volatile contents, there is still a paucity of work that analyses the association between hit-and-run impacts and the presence of moderately volatiles on Mercury's surface. 
Our work does not address this issue, although this is an important constraint that should be taken into account in future studies. Current studies argue that giant impacts would not necessarily result in moderately volatile depletion of proto-Mercury \citep{Hubbard2014,stewartetal16, ebel_stewart18}. 
But even if most of the volatile content was removed by a giant impact, Mercury could have experienced subsequent non-erosive impacts from comets or leftover planetesimals, which delivered volatile material to its surface \citep[][]{Morb12,Lawrence13, Brasseretal16, Chau18, Hyodoetal21}, or could have accreted interplanetary dust on longer timescales \citep[][]{frantsevaetal22}. 

The scenario proposed in this work occurs during the initial tens of millions of years of planet formation, when several mechanisms could prevent substantial debris reaccretion. The presence of planetary embryos and planetesimals would provide multiple gravitational interactions capable of scattering debris into different orbital regions. The proto-Mercury remnant itself could experience orbital evolution through these interactions, potentially moving inward due to dynamical friction and resonant interactions, creating separation from the debris field. Additionally, the remaining disk material could provide weak drag forces, affecting debris evolution according to size distribution. The combination of dynamical scattering and drag forces, coupled with the active environment of early planet formation, provides natural pathways for debris evolution that could prevent significant reaccretion while avoiding major perturbations to other growing planets.

Finally, it is not our intention with this work to claim that certain scenarios are incapable of reproducing the current characteristics of Mercury. Rather, we aimed to broaden the scope of plausible scenarios by presenting ones that are more frequent in numerical simulations, less constrained in planetary contexts, and thus more likely. Moreover, future research will explore these processes to determine under what conditions our results remain consistent with Mercury's internal structure while also agreeing with the structure of the inner solar system.

\section{Methods}\label{sec4}

\subsection{Smoothed Particle Hydrodynamics}

To model the collisions, we use the \texttt{miluphcuda}\footnote{The SPH code \texttt{miluphcuda} is in active development and publicly available at \url{github.com/christophmschaefer/miluphcuda}.} Smoothed Particle Hydrodynamics (SPH) code \citep{Shaferetal2016, SCHAFER20}. The bodies are assumed to be fully differentiated and composed of an iron core and a basalt mantle. We use the M-ANEOS equations of state \citep[][]{Thompson1990,melosh2007} to model the thermodynamic response of materials, which is widely used for giant impact simulations \citep[e.g.][]{Benz88,Benz2007,Asphaug_Reufer_14,saridetal14,Chau18}. Towards a more realistic material description (compared to simple hydrodynamical bodies), we include full material strength modelling by the Grady-Kipp fragmentation model for the basalt mantle \citep{gradykipp1980,benz1995} (with Weibull constants $k=5\times10^{34}$ and $m=8.5$ \cite{melosh1989}) combined with a von Mises yield limit to simulate plastic deformation for all materials (with yield limits of 10.5\,GPa for iron and 3.5\,GPa for basalt \cite{melosh1989}). Given the uncertainties about parameter values and even within strength models, it remains extremely difficult to accurately estimate the influence of different models in our simulations. Previous works have shown that material strength can significantly influence collision outcomes up to approximately Mars-sized bodies \cite{Jutzi2015,burgershafer17,Emsenhuberetal2018}.

We assume that the projectile is composed of a 70 wt\% silicate (basalt) mantle and a 30 wt\% iron core \citep[e.g.][]{Benz2007,Asphaug_Reufer_14,saridetal14,Chau18}. We consider two possible configurations for the target: one with a 70 wt\% rocky mantle and a 30 wt\% iron core, and another with a 50 wt\% rocky mantle and a 50 wt\% iron core \citep[e.g.][]{Emsenhuberetal2018}, aiming to explore possibilities with extreme planetary compositions. 

In order to obtain sufficiently equilibrated and physical initial configurations, we apply a semi-analytical relaxation technique, where the hydrostatic radial structures for the main quantities (density, pressure, internal energy) are computed and applied to the SPH particle setup \citep{Diehletal12, burgeretal18}. This method saves computing time by eliminating the need for further numerical relaxation while still producing low-noise initial conditions that are very close to equilibrium.

The simulations use an intermediate resolution with a total of $10^5$ SPH particles, distributed between the target ($\sim60$k to 80k particles) and the projectile (proto-Mercury, $\sim20$k to 40k particles). The overall simulation time span is approximately 2 days \citep[e.g.][]{Benz2007,Asphaug_Reufer_14,saridetal14,Chau18}, when the final masses have sufficiently converged. We also performed some simulations with higher resolution ($0.5\times10^{6}$ and $10^{6}$ SPH particles), and extended some simulations for up to 15 days (1/6 of Mercury's period).

\subsection{Impact Configurations}

Figure \ref{histo} compiles the distribution of hit-and-run collisions registered in the simulations by \cite{francoetal22}, where we first considered all collisions in a hit-and-run regime in which the projectile has at least one Mercury mass (grey histograms). From these collisions, we considered those in which the mass of the projectile is close to 2.36 Mercury mass (blue histograms), and the histograms in red represent the collisions that yield a second largest remnant with a mass equal to one current mass of Mercury according to the scaling laws adopted here.

The collision configurations that we use in our simulations initially depend on the following parameters: (i) the masses of the target $M_t$, and the projectile $M_p$, (ii) the impact angle $\theta$, and (iii) the relative velocity between the two bodies $v$. We are interested in grazing collisions that lead to the projectile's mantle stripping without mutual accretion of the bodies; i.e., all configurations correspond to a hit-and-run impact regime \citep[][]{LS12, Asphaug_Reufer_14, Chau18}.

In the simulations, the projectile mass is set to $0.13\,M_{\oplus}$, which is more than twice Mercury's current mass ($M_{\mercury} = 0.055\,M_{\oplus}$). The silicate-to-iron ratio of 70:30 ensures enough \textcolor{blue}{rocky} abundance and a comparable average composition to other terrestrial planets \cite{Cameronetal1988,Asphaug_Agnor_04,Asphaug_Reufer_14,saridetal14,Chau18}. The target mass is set to be between $\sim1.5$ and 4.5 times the projectile mass, specifically in the range of 0.2 $M_{\oplus}$ to 0.6 $M_{\oplus}$.

Impact angles and relative velocities are linked by scaling laws \citep{LS12} to produce the largest projectile remnant with the current mass of Mercury. According to these laws, in principle, several possible configurations are able to produce the same result, as shown in Fig. \ref{fig_vi_thetai}. The curves in that figure represent the relative velocity, $v$, scaled to the mutual escape velocity, $v_\mathrm{esc}$, as a function of the impact angle, $\theta$, for different masses of the target. The mutual escape velocity is given by:
\begin{equation}
    v_{\mathrm{esc}}=\sqrt{\frac{2G(M_t+M_p)}{R_t+R_p}},
    \label{equation_vesc}
\end{equation}
where $R_t$, $R_p$ are the radii of the target and the projectile, respectively.

First we set the desired largest remnant mass and calculate the required impact energy with the following the universal law for the mass of the largest remnant \citep{LS12, SL12}:
\begin{align}
    M_{\mathrm{lr}} &= \bigg(1-\frac{Q_R}{2Q'^*_{RD}}\bigg)M_{\mathrm{tot}},
    \label{eq:mf2}
\end{align}

where $Q_R/Q'^*_{RD}$ is the required impact energy scaled by the catastrophic disruption criteria, i.e., the specific impact energy needed to disperse half of the colliding total mass, $M_{\mathrm{tot}}$. We then compute the corresponding impact velocity for each impact angle using the following equation:

\begin{align}
    V_{\mathrm{lr}} &= \sqrt{\frac{2Q_R M_{\mathrm{tot}}}{\mu}},
    \label{eq:vimp}
\end{align}

where $\mu = M_pM_t/M_{\mathrm{tot}}$ is the reduced mass. We considered two different sets of impact configurations. The first set corresponds to the minimum values for impact velocities predicted by the scaling laws, represented by the black dots in Fig. \ref{fig_vi_thetai}. We call these the \textit{critical angles} set, because with such impact angles the collisions require the lowest impact energy. In such configurations, the impact angles are in the range of $40\degree$ to $65\degree$, and the relative velocities are in the range of $2.7\,v_{\mathrm{esc}}$ to $3.8\,v_{\mathrm{esc}}$. 

The second set of configurations corresponds to impact angles that are around 30\% to 40\% smaller than the critical angles in set A. These values are represented by the blue crossed circles in Fig. \ref{fig_vi_thetai}, and lie in the range of $33\degree$ to $41.5\degree$, with relative velocities in the range of $3\,v_{\mathrm{esc}}$ to $4\,v_{\mathrm{esc}}$. We call this the \textit{decreased angles} set. These configurations allow for a larger cross section between the colliding bodies and, consequently, a longer interacting length $L_{\mathrm{int}}$. This parameter represents the projected length of the interacting mass of the projectile onto the target \citep[e.g.][]{LS12,SL12,Qui16} and is given by
\begin{equation}
    L_{\mathrm{int}} = (R_t+R_p)(1-\sin\theta),
\end{equation}
for a grazing collision, when:
\begin{equation}
    \sin\theta > 1-\frac{2R_p}{R_t+R_p}.
\end{equation}
It is schematically represented in Fig. \ref{col_scheme}.

We do not consider impact angles greater than the critical ones -- in particular, those having the same impact velocities of the decreased angles -- because larger impact angles will decrease the cross section between the bodies, and such grazing impacts do not effectively strip the projectile mantle at the corresponding impact velocities \citep{Chau18}.

We also do not consider initial rotation, neither in the target nor in the projectile, because the spin velocities of the colliding bodies (typically $\sim1~\mathrm{km\,s}^{-1}$ on the surface) would be much lower than the impact velocities involved in the collisions ($\sim20~\mathrm{km\,s}^{-1}$). If both bodies were spinning, we may assume that there will be a small increase (if both are prograde) or reduction (if one is prograde and the other retrograde) in mantle stripping, causing the same effect as slightly increasing or decreasing the impact velocity for non-rotating bodies. We conclude that this effect would not be sufficient to significantly alter the overall results of our simulations, and thus we may disregard the bodies' spins. However, it should be noted that spinning generally tends to slightly increase mantle stripping \citep[e.g.][]{Chau18}. The impact configurations used in our simulations are listed in Table \ref{configurations}. 

\subsection{Treatment of Collision Outcomes}

At the end of each simulation, we determine the groups of SPH particles that represent the two major remnant bodies. First, we consider the centroid of different particle clusters and calculate where the majority of the particles are concentrated, a.k.a. the clump. Following that, we check whether the relative velocities of surrounding particles are below the clump's escape velocity, and also whether those particles are within a restricted sphere of influence of $5\times10^{-2}$ Hill radii of the clump, relative to the post-collision remnant of the target. This approach is similar to the one used by SKID (Spline Kernel Interpolative Denmax) \cite[][]{Stadel2001}, a friends-of-friends algorithm that determines whether a given SPH particle is part of a gravitationally bound clump \citep[see][for details]{Chau18,Benincasa2019, Timpe2020}.

\subsection{Dependence on SPH Resolution}

\begin{figure}[h]%
    \centering
    \includegraphics[width=\textwidth]{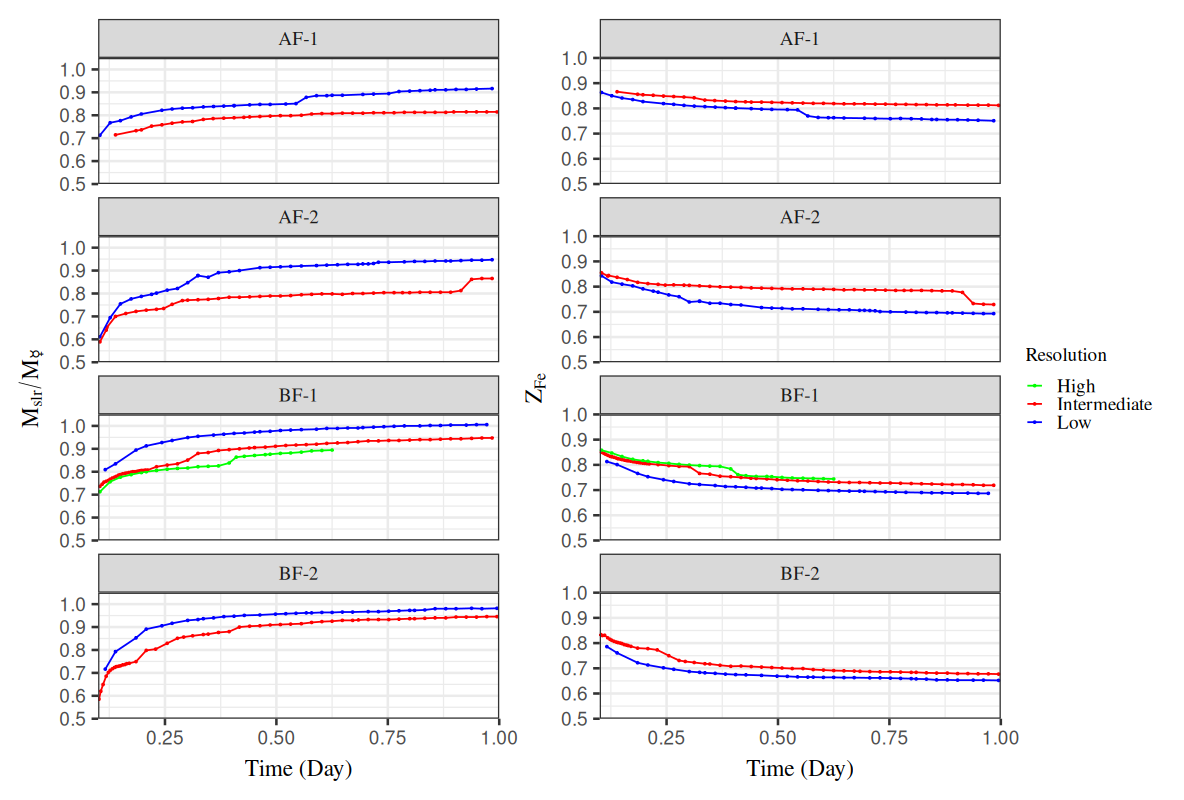}
    \caption{Resolution comparison of our best simulations in terms of scaled mass (left) and iron-mass fraction (right) of the second largest remnant over time. AF and BF (-1) and (-2) correspond to simulations involving targets of 0.2 and 0.6 Earth masses, respectively. Low, intermediate, and high resolutions correspond to $1\times10^5$, $5\times10^5$, and $1\times10^6$ total particles, respectively.}
    \label{resolutions}
\end{figure}

The resolution of the simulation is another determining factor in the outcomes. In our simulations, we used a low resolution of $10^5$ particles. Then we ran higher-resolution simulations with $0.5\times10^6$ and $10^6$ particles throughout the system. As we can see in Figure \ref{resolutions}, the resolution has an effect on the results considering the first 24 hours of the simulations. However, in our case (hit-and-run giant impact), the change is negligible, and this may become even more apparent as the system evolves over time. Changes in the iron-mass fraction results are even smaller depending on the resolution.

The most significant differences in the mass and iron-mass fraction in higher-resolution simulations (AF) compared to the lower ones are related to the presence of other small individual clusters that are not accreted to the main cluster. They evolve around the projectile remnant and are not observed in most low-resolution and BF high-resolution simulations.

These several clusters are small aggregates of silicate ejecta that promote a delay in the main cluster's accretion process. As a result, the accretion time scale of the remnant increases. Because they are within a few Hill radii of influence of the main body and have similar relative velocities, it is very likely that these small clusters will eventually be reaccreted later to the main body. In the end, they will increase the final remnant mass while decreasing the final iron mass fraction, yielding results similar to lower-resolution simulations. Although we have observed that the mass of the remnant decreases while the final fraction of iron mass increases in comparison to the original simulations when these structures are still present during the impact evolution. Hence, these results show that our simulations are consistent and that increasing system resolution has negligible effects on the outcomes.

\subsection{Dependence on simulation time span}
\label{sec:time}
In the simulations that produce the best results, the projectile has its mantle almost completely stripped out and, subsequently, begins to reaccrete the material around it (cf. Fig. \ref{snap}). After two days of evolution, the remnant is predominantly made up of iron material and its mass no longer increases significantly. The rocky material reaccreted by the remnant comes almost completely from the original projectile, and less than 0.5\% of the reaccreted material comes from the original target. This is because the projectile's velocity is not reduced abruptly following the collision, and most of the lost mantle material remains around the impact zone. The fraction of the projectile's mantle material that is unaffected by the target and eventually unbinds the projectile's core maintains the same velocity as the iron aggregate, regrouping and incorporating rocky content into the final remnant.

Most of the ejecta are reaccreted to the target because it is the largest body in the configuration. However, at the end of our simulations, there is still about 20\% of the original rocky ejecta that has not been accreted to either the target or the projectile. To achieve a successful result, the projectile remnant must avoid reassembling all its lost material. To assess this, we extended our best simulations of each setup group (AF and BF) to 15 days. We observed that our results did not change significantly after that time. Compared to results after only 2 days, the projectile remnant had a slight increase in the final masses (2-5\%) and a decrease in the iron-mass fractions (1-3\%).

\backmatter

\bmhead{Acknowledgments}
P.~Franco, F.~Roig, and O.~Winter acknowledge support from the Brazilian National Council of Research - CNPq (processes 305210/2018-1 and 306009/2019-6).
C.~Burger and C.~M.~Sch{\"a}fer appreciate support by the German Research Foundation - DFG (project 285676328).
Simulations have been performed using the cluster of the Grupo de Dinâmica Orbital e Planetologia of UNESP, financed by the São Paulo State Research Foundation - FAPESP (process 2016/24561-0).
The authors also thank the Brazilian Federal Agency for Support and Evaluation of Graduate Education - CAPES, in the scope of the Program CAPES-PrInt, process 88887.310463/2018-00, International Cooperation Project number 3266. 

\section*{Declarations}
\begin{itemize}
\item Funding: Not applicable
\item Conflict of interest/Competing interests: The author(s) declare no competing interests.
\item Ethics approval: Not applicable.
\item Consent to participate: Not applicable.
\item Consent for publication: Not applicable.
\item Availability of data and materials: The datasets generated during and/or analysed during the current study are available from the corresponding author on reasonable request.
\item Code availability: The SPH code \texttt{miluphcuda} used in this paper is publicly available at \url{github.com/christophmschaefer/miluphcuda}. GNU GENERAL PUBLIC LICENSE Version 3, 29 June 2007. Copyright (C) 2007 Free Software Foundation, Inc. https://fsf.org/.
Everyone is permitted to copy and distribute verbatim copies of this license document, but changing it is not allowed.
\item Authors' contributions: Not applicable.
\end{itemize}


\goodbreak
\bibliography{mercury-bibliography}  

\newpage
\begin{table}[h]
    \caption{Impact configurations considered in our simulations.}
    \label{configurations}
    \resizebox{\textwidth}{!}{  
    \begin{tabular}{ccccccc}
        \hline
        \\
        Base setup & $M_t$ ($M_{\oplus}$) & $\rho_t$ ($\mathrm{g\,cm}^{-3}$) & $\theta_i$ ($\degree$)  & $v_i/v_{\mathrm{esc}}$ & $Z_{\text{Fe,slr}}$ & $M_{\mathrm{slr}}$ ($M_{\oplus}$) \\ \\ 
        \hline
        \\
        \multirow{5}{*}{AC} & 0.2 & 5.17 & 48.8 & 3.71 & 0.36 & 0.108 \\ 
           & 0.3 & 5.44 & 55.1 & 3.35 & 0.33 & 0.117 \\ 
           & 0.4 & 5.65 & 58.9 & 3.09 & 0.32 & 0.121 \\ 
           & 0.5 & 5.82 & 61.2 & 2.89 & 0.32 & 0.123 \\
           & 0.6 & 5.98 & 63.5 & 2.72 & 0.32 & 0.124 \\ \\
        \multirow{5}{*}{AD} & 0.2 & 5.17 & 34.5 & 3.76 & 0.63 & 0.062 \\ 
           & 0.3 & 5.44 & 38.0 & 3.45 & 0.60 & 0.065 \\ 
           & 0.4 & 5.65 & 39.0 & 3.24 & 0.63 & 0.062 \\ 
           & 0.5 & 5.82 & 40.0 & 3.08 & 0.65 & 0.059 \\ 
           & 0.6 & 5.98 & 41.5 & 2.94 & 0.64 & 0.060 \\ \\
        \multirow{2}{*}{AF} & 0.2 & 5.17 & 31.6 & 3.78 & 0.73 & 0.052 \\
           & 0.6 & 5.98 & 40.3 & 2.96 & 0.69 & 0.053 \\ \\
        \hline
        \\
        \multirow{2}{*}{BC} & 0.2 & 4.41 & 42.5 & 3.76 & 0.43 & 0.090 \\
           & 0.6 & 5.09 & 60.0 & 2.77 & 0.33 & 0.117 \\ \\
        \multirow{2}{*}{BD} & 0.2 & 4.41 & 35.3 & 3.77 & 0.60 & 0.065 \\
           & 0.6 & 5.09 & 42.0 & 2.90 & 0.61 & 0.061 \\ \\
        \multirow{2}{*}{BF} & 0.2 & 4.41 & 32.5 & 3.78 & 0.68 & 0.056 \\
           & 0.6 & 5.09 & 41.0 & 2.91 & 0.65 & 0.055 \\ \\
        \hline
        \\
    \end{tabular}
    }
    Notes: The base setup is coded with two letters in the first column. The first letter refers to the initial core-mass ratio, $Z_\mathrm{Fe}$, of the target: letter A corresponds to $Z_\mathrm{Fe}=0.5$ and letter B corresponds to $Z_\mathrm{Fe}=0.3$. The second letter refers to the impact angles setup: C is for the critical angles, D is for the decreased angles, and F is for the fine-tuned angles (see Online Methods). M$_t$ and $\rho_t$ are the initial mass and mean density of the target, $\theta_i$ is the impact angle, $v_i/v_{\mathrm{esc}}$ is the scaled impact velocity,  $Z_{\mathrm{Fe,slr}}$ and $M_{\mathrm{slr}}$ are the final core-mass ratio and mass of the second largest remnant, i.e. the Mercury candidate.
\end{table}

\newpage

\newpage

\newpage

\newpage

\newpage

\newpage

\end{document}